\documentclass[preprint,amssymb, aps, prl,floatfix,superscriptaddress]{revtex4-1}
\usepackage{graphicx}
\usepackage{bm}
\renewcommand{\vec}[1]{\bm{#1}}
\newcommand{\FPMvB}{FPMVB}
\begin{document}

\title{Calculating the Magnetic Anisotropy of Rare-Earth---Transition-Metal Ferrimagnets}

\author{Christopher E.\ Patrick}
\email{c.patrick.1@warwick.ac.uk}
\author{Santosh Kumar}
\author{Geetha Balakrishnan}
\author{Rachel S.\ Edwards}
\author{Martin R.\ Lees}
\affiliation{Department of Physics, University of Warwick, Coventry CV4 7AL, United Kingdom}
\author{Leon Petit}
\affiliation{Daresbury Laboratory, Daresbury, Warrington WA4 4AD, United Kingdom}
\author{Julie B.\ Staunton}
\affiliation{Department of Physics, University of Warwick, Coventry CV4 7AL, United Kingdom}
\date{\today}

\begin{abstract}
Magnetocrystalline anisotropy, the microscopic origin of permanent magnetism, 
is often explained in terms of ferromagnets.
However, the best performing
permanent magnets based on rare earths and transition metals (RE-TM)
are in fact \emph{ferri}magnets, consisting of a number of magnetic 
sublattices. 
Here we show how a naive calculation of the magnetocrystalline anisotropy
of the classic RE-TM ferrimagnet GdCo$_5$ gives numbers which are too large
at 0~K and exhibit the wrong temperature dependence.
We solve this problem by introducing a first-principles approach to calculate temperature-dependent 
magnetization vs.\ field (FPMVB) curves, mirroring the experiments
actually used to determine the anisotropy.
We pair our calculations with measurements on a recently-grown
single crystal of GdCo$_5$, and find excellent agreement.
The FPMVB \ approach demonstrates a new level of sophistication
in the use of first-principles calculations to understand RE-TM magnets.
\end{abstract}
\maketitle
High-performance permanent magnets, as found in
generators, sensors and actuators,
are characterized by a large volume magnetization
and a high coercivity~\cite{Chikazumibook}.
The coercivity --- which measures the
resistance to demagnetization by external fields ---
is upper-bounded by the material's magnetic anisotropy~\cite{Kronmuller1987},
which in qualitative terms describes a preference 
for magnetization in particular directions.
Magnetic anisotropy may be partitioned into
two contributions: the shape anisotropy, determined
by the macroscopic dimensions of the sample,
and the magnetocrystalline anisotropy (MCA), which
depends only on the material's crystal structure and chemical
composition.
Horseshoe magnets provide a practical demonstration of shape anisotropy,
but the MCA is less intuitive,
arising from the relativistic quantum mechanical
coupling of spin and orbital degrees of 
freedom~\cite{Strangebook}.

Permanent magnet technology was revolutionized with
the discovery of the rare-earth/transition-metal (RE-TM)
magnet class, beginning with Sm-Co magnets in 1967~\cite{Strnat1967}
(whose high-temperature performance is still unmatched~\cite{Gutfleisch2011}),
followed by the world-leading workhorse magnets based on 
Nd-Fe-B~\cite{Sagawa1984,Croat1984}.
With the TM providing the large volume magnetization,
careful choice of RE yields MCA values
which massively exceed the shape anisotropy contribution~\cite{Coey2011}.
RE-TM magnets are now indispensable to
everyday life, but their significant economic and environmental cost 
has inspired a global research effort aimed at replacing 
the critical materials required in their manufacture~\cite{Skomski2013}.

In order to perform a targeted search for new materials
it is necessary to fully understand the huge MCA of existing RE-TM magnets.
An impressive body of theoretical work based on crystal
field theory has been built up over decades~\cite{Kuzmin2008},
where model parameters are determined
from experiment (e.g.\ Ref.~\cite{Tiesong1991}) or
electronic structure calculations~\cite{Richter1998,Kuzmin2004,Delange2017}.
An alternative and increasingly more common approach
is to use these electronic structure calculations,
usually based on density-functional theory (DFT),
to calculate the material's magnetic properties directly
without recourse to the crystal field 
picture~\cite{Steinbeck2001,Larson2004,Pang2009,Matsumoto2014,Landa2017}. 

Calculating the MCA of RE-TM magnets 
presents a number of challenges to electronic structure theory.
The interaction of localized RE-4$f$ electrons with their
itinerant TM counterparts is poorly described within the most 
widely-used first-principles
methodology, the local spin-density approximation (LSDA)~\cite{Richter1998}.
Indeed, the MCA is inextricably linked to orbital magnetism whose
contribution to the exchange-correlation energy is missing in spin-only 
DFT~\cite{Eriksson19901,Eschrig2005}.
MCA energies are generally a few meV per formula unit, necessitating
a very high degree of numerical convergence~\cite{Daalderop1996}.
Finally, the MCA depends strongly on temperature, so a practical
theory of RE-TM magnets must go beyond zero-temperature DFT 
and include thermal disorder~\cite{Staunton2004}.

Even when these significant challenges have been overcome,
there is a more fundamental problem.
Experiments access the MCA indirectly, measuring the
change in magnetization of a material when an external field is applied
in different directions.
By contrast, calculations usually access the MCA directly by
evaluating the change in energy when the material is magnetized in different
directions, with no reference to an external field.
These experimental and computational approaches arrive at the same MCA energy
provided one is studying a \emph{ferro}magnet.
However, the majority of RE-TM magnets (and many other
technologically-important magnetic materials)
are \emph{ferri}magnets, i.e.\ they are composed 
of sublattices with magnetic
moments of distinct magnitudes and orientations.
Crucially the application of an external field
may introduce canting between these sublattices,
affecting the measured magnetization.
Thus the standard theoretical approach of 
ignoring the external field is hard to reconcile with
real experiments on ferrimagnets.

In this Letter, through a combination of 
calculations and experiments, we provide the hitherto missing link 
between electronic structure theory and practical measurements of the MCA.
Specifically, we show how to directly simulate experiments
by calculating, from first principles (FP), how the measured magnetization ($M$)
varies as a function of field ($B$) applied along different
directions and at different temperatures.
We apply our ``\FPMvB'' approach to
the RE-TM ferro and ferrimagnets YCo$_5$ and GdCo$_5$,
which are isostructural to the technologically-important
SmCo$_5$~\cite{Kumar1988} and, in the case of GdCo$_5$, a source of controversy 
in the literature~\cite{Buschow1974,Ermolenko1976,Rinaldi1979,Yermolenko1980,Ballou1986,
Radwanski1986,Ballou1987,Gerard1992,Radwanski1992,Franse1993,Zhao1991}.
Pairing \FPMvB~with new measurements of the MCA 
of GdCo$_5$ allows us to resolve this controversy.
More generally, \FPMvB \ enables a new level of collaboration
between theory and experiment in understanding
the magnetic anisotropy of ferrimagnetic materials.

The electronic structure theory behind \FPMvB~
treats magnetic disorder at a finite temperature $T$
within the disordered local moment (DLM) picture~\cite{Gyorffy1985,Staunton2006}.
The methodology allows the calculation of the magnetization of each sublattice $i$,
$\vec{M_i}(T) = M_i(T)\vec{\hat{M}_i}$, and the torque quantity $\partial F(T)/\partial\vec{\hat{M}_i}$,
where $F$ is an approximation to the temperature dependent free energy.
$\partial F(T)/\partial\vec{\hat{M}_i}$ accounts for the anisotropy arising from
the spin-orbit interaction, while the contribution from the classical magnetic dipole 
interaction is computed numerically~\footnote{We performed
a sum over dipoles~\cite{Chikazumibook} using the calculated
$\vec{M_i}(T)$ out to a radius of 20~nm.}.
Many of the technical details of the DFT-DLM calculations~\cite{Bruno1997,
Gyorffy1985,Strange1984,Daene2009,Vosko1980,Lueders2005} were described
in our recent study of the magnetization of the same compounds~\cite{Patrick2017};
the extensions to calculate the torques
are described in Ref.~\cite{Staunton2006}.
The Gd-$4f$ electrons are treated with the local self-interaction 
correction~\cite{Lueders2005},
and we have also implemented the orbital polarization correction~\cite{Eriksson19901} 
following Refs.~\cite{Ebert1996,Ebertbook} 
using reported Racah parameters~\cite{Steinbeck20012}.
Details are given as Supplemental Material (SM)~\cite{SM}.

\begin{figure}
\includegraphics{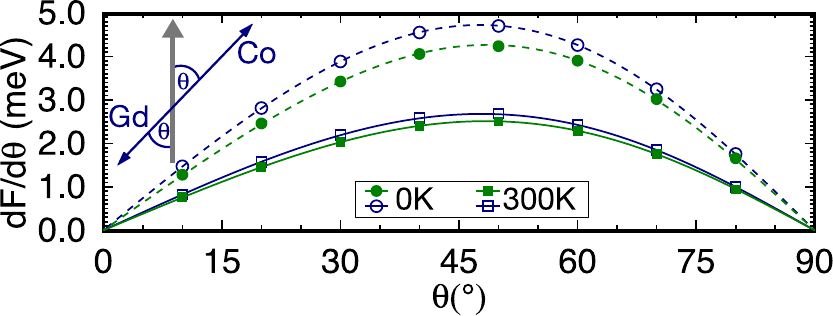}
\caption{
Data points and fits of $dF/d\theta$ calculated for GdCo$_5$ (blue, empty symbols;
Gd and Co moments held antiparallel)
and YCo$_5$ (green, filled symbols),
at 0~and~300~K.
\label{fig.Kvang}
}
\end{figure}
YCo$_5$ and GdCo$_5$ crystallize in the CaCu$_5$ structure,
consisting of alternating hexagonal RCo$_{2c}$/Co$_{3g}$ layers~\cite{Kumar1988}.
Y is nonmagnetic, while in GdCo$_5$ the large 
spin moment of Gd (originating mainly from its half-filled 4$f$ shell)
aligns antiferromagnetically with the Co moments.
We now consider a ``standard'' calculation of the MCA based on
a rigid rotation of the magnetization.
If the Gd and Co moments are held antiparallel, GdCo$_5$
is effectively a ferromagnet with reduced moment
$M_\mathrm{Co} - M_\mathrm{Gd}$.
Then, from the hexagonal symmetry we expect the angular dependence of the
free energy to follow
$\kappa_1 \sin^2 \theta + \kappa_2 \sin^4 \theta + \mathcal{O}(\sin^6 \theta)$,
where $\theta$ is the polar angle between the crystallographic $c$ axis 
and the magnetization direction.
The constants $\kappa_1,\kappa_2$ determine the change in free energy $\Delta F$,
calculated e.g.\ from the force theorem~\cite{Daalderop1990} or the torque $dF/d\theta$~\cite{Wang1996}.

In Fig.~\ref{fig.Kvang} we show $dF/d\theta$ 
calculated for ferromagnetic YCo$_5$ and GdCo$_5$ at 0~and~300~K.
Fitting the data to the derivative of the textbook expression,
$\sin 2\theta  (\kappa_1  + 2\kappa_2\sin^2\theta)$,
finds $\kappa_1$ and $\kappa_2$ to be positive (easy $c$ axis) with 
$\kappa_1$ an order of magnitude larger than $\kappa_2$.
Considering experimentally measured anisotropy constants in
the literature, for YCo$_5$
our $\kappa_1$ value of 3.67~meV  (all energies are per formula unit, f.u.) at 0~K 
compares favorably to the values of 3.6~and~3.9~meV reported
in Refs.~\cite{Yermolenko1980} and \cite{Alameda1981}.
At 300~K, our value of 2.19~meV exhibits a slightly faster decay with
temperature compared to experiment  (2.6~and~3.0~meV), which we attribute
to our use of a classical spin hamiltonian in
the DLM picture~\cite{Gyorffy1985,Patrick2017}.
However, for GdCo$_5$ our calculated values of $\kappa_1$  show very poor agreement
with experiments~\cite{Ermolenko1976,Ballou1986}.
First, at 0~K we find $\kappa_1$ to be larger than YCo$_5$ (4.26~meV),
while experimentally the anisotropy constant is much smaller (1.5, 2.1~meV).
Second, we find $\kappa_1$ decreases with temperature (2.39~meV at 300~K)
while experimentally the anisotropy constant \emph{increases} (2.7, 2.8~meV).

To understand these discrepancies we must ask how the anisotropy energies
were actually measured.
Torque magnetometry provides an accurate method of accessing the
MCA~\cite{Klein1975}, but is technically challenging in RE-TM magnets, which require very high fields
to reach saturation~\cite{Buschowbook}.
Singular point detection~\cite{Paoluzi1994} and ferromagnetic resonance~\cite{Wang2003}
has also been used to investigate the MCA of polycrystalline and thin-film
samples.
However, the most commonly-used method for RE-TM magnets, employed in 
Refs.~\cite{Ermolenko1976,Ballou1986}, is based on the seminal
1954 work by Sucksmith and Thompson~\cite{Sucksmith1954} on 
the anisotropy of hexagonal ferromagnets.
This work provides a relation 
between the measured magnetization
$M_{\mathrm{ab}}$ and field $B$ applied in the hard plane
in terms of $\kappa_1$, $\kappa_2$ 
and the easy axis magnetization $M_0$~\cite{Sucksmith1954,SM}:
\begin{equation}
(BM_0/2) / \left(M_\mathrm{ab}/M_0\right)  \equiv \eta
= \kappa_1 + 2\kappa_2 \left(M_{\mathrm{ab}}/M_0\right)^2.
\label{eq.ST}
\end{equation}
Further introducing $m=(M_{\mathrm{ab}}/M_0)$, equation~\ref{eq.ST} shows that 
a plot of $\eta$ against $m^2$ should yield a straight line with $\kappa_1$
as the intercept.
Even though this ``Sucksmith-Thompson method''
was derived for ferromagnets, the technical procedure of 
plotting $\eta$ against $m^2$ can
be performed also for ferrimagnets like GdCo$_5$~\cite{Ermolenko1976,Ballou1986}.
In this case, the quantity extracted from the intercept 
is an effective anisotropy constant $K_\mathrm{eff}$
so, unlike YCo$_5$, the anisotropy
constants reported in Refs.~\cite{Ermolenko1976,Ballou1986}
are distinct from the $\kappa_1$ values extracted from Fig.~\ref{fig.Kvang}.
As recognized at the time of the original experiments~\cite{Rinaldi1979,
Yermolenko1980,Ballou1986,Radwanski1986}, the reduced value of 
$K_\mathrm{eff}$ with respect to $\kappa_1$ of YCo$_5$ is a fingerprint 
of canting between the Gd and Co sublattices.

Making contact with previous experiments thus requires
we obtain $K_\mathrm{eff}$.
To this end we have developed a scheme of calculating first-principles
hard-plane magnetization vs.\ field (\FPMvB) curves,
on which we perform the Sucksmith-Thompson analysis to directly
mirror the experiments.
The central concept of \FPMvB~is that at equilibrium, the torques from
the exchange, spin-orbit and dipole interactions must balance those arising
from the external field.
Then,
\begin{equation}
B = \frac{\partial F(T)}{\partial \theta_i}\frac{1}{M_i \cos \theta_i + \sum_j \sin\theta_j\frac{ \partial M_j}{ \partial \theta_i}}.
\label{eq.torquematch}	
\end{equation}
The magnetization at a given $B,T$ is determined by the angle set 
$\{\theta_\mathrm{Gd}, \theta_{\mathrm{Co_1}}, \theta_{\mathrm{Co_2}},...\}$
which satisfies equation~\ref{eq.torquematch} for every magnetic sublattice.
The spin-orbit interaction breaks the symmetry of the Co$_{3g}$  atoms
such that altogether there are four independent angles to vary for GdCo$_5$.
The second term in the denominator of equation~\ref{eq.torquematch}
reflects that the magnetic moments themselves might depend on $\theta_i$
(magnetization anisotropy).
We have tested (i) neglecting this contribution and (ii) modeling the
dependence as $M_i(\theta_i) = M_{0i}(1-p_i \sin^2\theta_i)$, where $M_{0i}$ and
$p_i$ are parameterized from our calculations.

\begin{figure}
\includegraphics{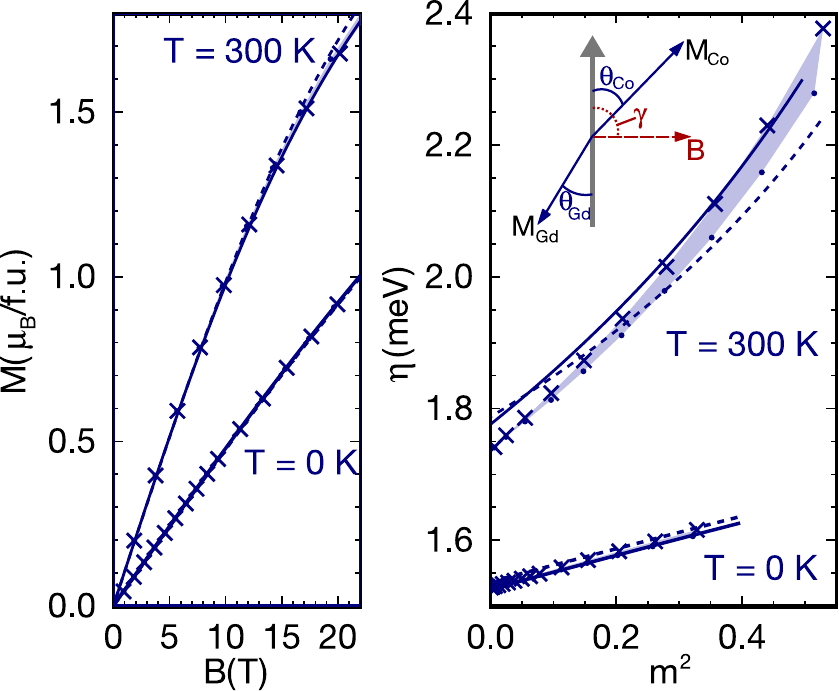}
\caption{
Magnetization of GdCo$_5$ vs.\ applied magnetic field shown on a standard plot (left panel)
or after the Sucksmith-Thompson analysis (eq.~\ref{eq.ST}, right panel).
Crosses/circles are calculated with methods (i)/(ii) discussed in the text,
and the area between them shaded as a guide to the eye.
Note the two methods are effectively indistinguishable in the left panel.
The dashed/solid lines are calculated from the model free energies $F_1$
and $F_2$.
The right panel also shows the geometry of the magnetization and field
with respect to the crystallographic $c$-axis (thick gray arrow).
\label{fig.MvB}
}
\end{figure}
Figure~\ref{fig.MvB} shows \FPMvB~curves
of GdCo$_5$ calculated using equation~\ref{eq.torquematch} with
methods (i) and (ii), (crosses and circles) which yield virtually identical values of $K_\mathrm{eff}$.
The $M$~vs.~$B$ curves in the left panel resemble those of
a ferromagnet where, as the temperature increases, 
it becomes easier to rotate the moments away from the easy axis
so that a given $B$ field induces a larger magnetization.
However, plotting $\eta$ against $m^2$ in the right panel tells a more
interesting story.
The effective anisotropy constant $K_\mathrm{eff}$ ($y$-axis intercept)
at 0~K is 1.53~meV, much smaller than $\kappa_1$ of YCo$_5$.
Furthermore $K_\mathrm{eff}$ increases with temperature, to 1.74~meV at 300~K.
Therefore, in contrast to the standard calculations of Fig.~\ref{fig.Kvang},
the \FPMvB~approach reproduces the
experimental behavior of Refs.~\cite{Ermolenko1976,Ballou1986}.

Our \FPMvB~calculations provide a microscopic insight into the magnetization process.
For instance at 0~K and 9~T, we calculate that the cobalt
moments rotate away from the easy axis by 6.1$^\circ$.
By contrast the Gd moments have rotated by only 3.9$^\circ$, i.e.\ the ideal
180$^\circ$ Gd-Co alignment has reduced by 2.2$^\circ$ (the geometry is shown 
in Fig.~\ref{fig.MvB}).
We also find canting between the different Co sublattices, but not by
more than 0.1$^\circ$ at both 0~and~300~K (the calculated angles as a function
of field are shown in the SM~\cite{SM}).
This Co-Co canting is small  
thanks to the Co-Co ferromagnetic exchange 
interaction, which remains strong over a wide temperature range~\cite{Patrick2017}.
The temperature dependence of $K_\mathrm{eff}$ can be traced to the fact
that the easy axis magnetization $M_0$ of GdCo$_5$ initially increases 
with temperature~\cite{Patrick2017}.
Even if $M_{\mathrm{ab}}$ increases with temperature at a given field, a faster increase
in $M_0$ can lead to an overall hardening in $K_\mathrm{eff}$ (equation~\ref{eq.ST}).

We assign the canting in GdCo$_5$ to a
delicate competition between the exchange
interaction favoring antiparallel Co/Gd moments,
uniaxial anisotropy favoring $c$-axis (anti)alignment,
and the external field trying
to rotate all moments into the hard plane.
We can quantify these interactions by looking for a model
parameterization of the free energy $F$.
Crucially we can train the model with an arbitrarily large 
set of first-principles calculations exploring sublattice orientations not accessible experimentally,
and test its performance against the torque calculations of equation~\ref{eq.torquematch}.
Neglecting the 0.1$^\circ$ canting within the cobalt sublattices
gives two free angles, $\theta_\mathrm{Gd}$ and $\theta_\mathrm{Co}$.
Including Gd-Co exchange $A$, uniaxial Co anisotropy $K_\mathrm{1,Co}$ and a dipolar 
contribution $S(\theta_\mathrm{Gd},\theta_\mathrm{Co})$~\cite{Ballou1987,SM}
leads naturally to a two-sublattice model~\cite{Radwanski1986},
\begin{eqnarray}
F_{1}(\theta_\mathrm{Gd},\theta_\mathrm{Co}) &=& -A\cos(\theta_\mathrm{Gd}-\theta_\mathrm{Co})
+ K_\mathrm{1,Co} \sin^2 \theta_\mathrm{Co}  \nonumber \\
&& + S(\theta_\mathrm{Gd},\theta_\mathrm{Co}).
\end{eqnarray}
The training calculations showed additional angular
dependences not captured by $F_1$, so we also investigated:
\begin{eqnarray}
F_{2}(\theta_\mathrm{Gd},\theta_\mathrm{Co}) &=& 
F_{1}(\theta_\mathrm{Gd},\theta_\mathrm{Co}) + K_\mathrm{2,Co} \sin^4 \theta_\mathrm{Co} \nonumber \\
&&+ K_\mathrm{1,Gd} \sin^2 \theta_\mathrm{Gd}.
\label{eq.F2}
\end{eqnarray}
As discussed below the training calculations showed no strong evidence 
of Gd-Co exchange anisotropy~\cite{Ballou1987,Gerard1992,Radwanski1992,Franse1993}.

The dashed (solid) lines in Fig.~\ref{fig.MvB} are the calculated $M$ vs.\ $B$ curves
obtained by minimizing $F_{1(2)} - \sum_i \vec{M_i}\cdot\vec{B}$.
The second term includes magnetization anisotropy on the cobalt moments~\cite{SM,Alameda1980}.
On the scale of the left panel both $F_1$ and $F_2$ give excellent fits to the torque
calculations, especially up to moderate fields.
The plot of $\eta$ against $m^2$ reveals some differences with
$F_2$ giving a marginally improved description of the data, but
$F_1$ already captures the most important physics.

\begin{figure}
\includegraphics{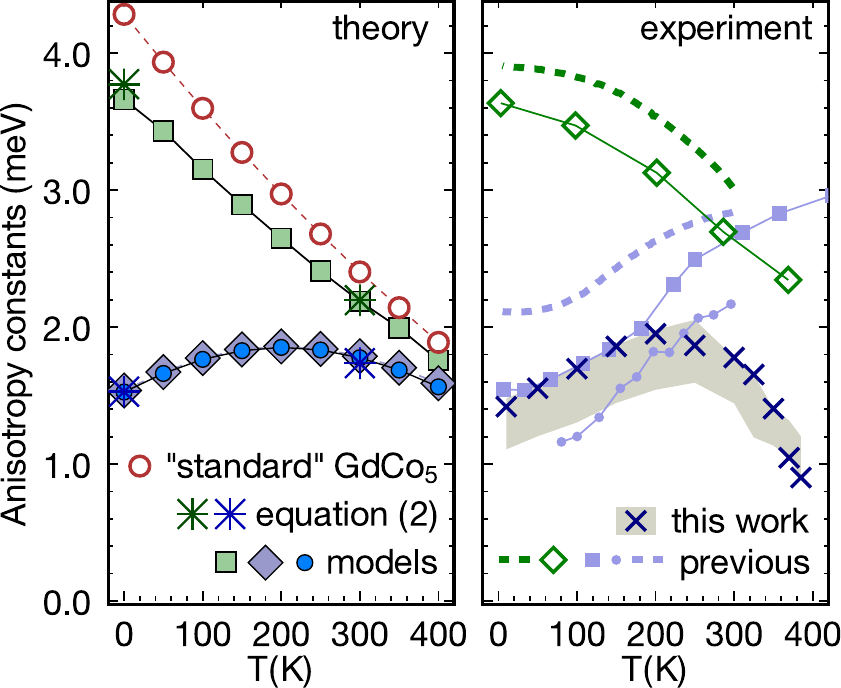}
\caption{
Anisotropy constants $K_\mathrm{eff}$ vs.\ temperature of YCo$_5$ (green) and GdCo$_5$ (blue).  
The left panel shows calculations using equation~\ref{eq.torquematch} 
at 0 and 300~K (stars),
or using parameterized model expressions $F_1$ (diamonds) and $F_2$ (circles),
and from Ref.~\cite{Alameda1980} (YCo$_5$, squares).
For GdCo$_5$ we also show in red $\kappa_1$ extracted from ``standard'' calculations where
the Gd and Co moments were held rigidly antiparallel (cf.\ Fig.~\ref{fig.Kvang}).
The experimental data in the right panel was measured by us for GdCo$_5$ (crosses,
with shaded background) or taken from  Refs.~\cite{Ermolenko1976}, \cite{Ballou1986} and 
\cite{Katayama1976} (squares, dashed lines, circles) and
Refs~\cite{Yermolenko1980}~and~\cite{Alameda1981} 
(green diamonds and dashed lines, YCo$_5$).
\label{fig.KvT}
}
\end{figure}
We also applied the \FPMvB~approach to YCo$_5$,  using equation~\ref{eq.torquematch}
and the model for $F$ introduced in Ref.~\cite{Alameda1980}.
Then, parameterizing the models~\cite{SM} over
the temperature range 0--400~K, 
calculating $M$ vs.\ $B$ curves and extracting
$K_\mathrm{eff}$ using the Sucksmith-Thompson plots gives
the results shown in the left panel of Fig.~\ref{fig.KvT}.
We also show $\kappa_1$ of GdCo$_5$
to emphasize the difference between \FPMvB~calculations 
and the ``standard'' ones of Fig.~\ref{fig.Kvang}.

Comparing $K_\mathrm{eff}$ to previously-published experimental
measurements on GdCo$_5$ raises some issues.
First, the three studies in the literature
report anisotropy constants which
differ by as much as 1~meV~\cite{Ermolenko1976,Ballou1986,Katayama1976}.
Indeed there was controversy over whether the observed
results were evidence of an anisotropic exchange interaction
between Gd and Co~\cite{Ballou1987,Gerard1992} or an artefact
of poor sample stoichiometry~\cite{Radwanski1992,Franse1993}.
Furthermore the only study performed above room
temperature~\cite{Ermolenko1976} reports without comment some peculiar behavior
where $K_\mathrm{eff}$ of GdCo$_5$ exceeds that of YCo$_5$
at high temperature~\cite{Yermolenko1980}, despite
conventional wisdom that the half-filled $4f$ shell of Gd
does not contribute to the anisotropy.

Our calculations do in fact show an excess in the rigid-moment
anisotropy of GdCo$_5$ of 16\% at 0~K (Fig.~\ref{fig.Kvang})
compared to YCo$_5$.
The authors of Refs.~\cite{Ballou1986,Ballou1987} fitted their experimental
data with a much larger excess of 50\%, while
the high-field study of Ref.~\cite{Radwanski1992} found 
(11 $\pm$ 15)\%, with the authors of that work attributing
the difference to an improved sample stoichiometry~\cite{Franse1993}.
Our calculated excess at 0~K is formed from two major contributions: the 
dipole interaction energy, which accounts
for 0.31~meV/f.u., and $K_{1,\mathrm{Gd}}$ (equation~\ref{eq.F2})
which we found to be 24\% the size of $K_{1,\mathrm{Co}}$.
The nonzero value of $K_{1,\mathrm{Gd}}$ is due to the 5$d$ electrons,
whose presence is evident from the Gd magnetization (7.47$\mu_B$ at 0~K).
We did not find a significant contribution from anisotropic exchange,
which we tested in two ways:
first by attempting to fit a term 
$A(1 - p'\sin^2\theta_\mathrm{Co})\cos(\theta_\mathrm{Gd} - \theta_\mathrm{Co})$
to our training set of calculations, 
and also by computing Curie temperatures
with the (rigidly antiparallel) magnetization directed either along the $c$ or $a$
axes.
We found the magnitude of the anisotropy ($p'$) to be smaller than 0.5\%
and negative at 0~K, and to decrease in magnitude as the temperature is raised.
Consistently the Curie temperature was found to be only 1~K higher for $a$ axis
alignment, which we do not consider significant.

However, our calculations do not predict
the $K_\mathrm{eff}$ value of GdCo$_5$ to exceed YCo$_5$.
Indeed, in Fig.~\ref{fig.KvT} $\kappa_1$ of GdCo$_5$ approaches
that of YCo$_5$ at high temperatures, which is significant because 
$\kappa_1$ provides an upper bound for $K_\mathrm{eff}$~\cite{Gerard1992}.
To resolve this final puzzle
we performed our own measurements of $K_\mathrm{eff}$
on the single crystal whose growth we reported recently~\cite{Patrick2017}.
Hard and easy axis magnetization curves up to 7~T were measured 
in a Quantum Design superconducting quantum interference 
device (SQUID) magnetometer, and the anisotropy constants extracted
from Sucksmith-Thompson plots~\cite{SM}.
The right panel of Fig.~\ref{fig.KvT} shows our newly measured data
as crosses.
Previously reported measurements are shown in faint blue/green for 
GdCo$_5$~\cite{Ermolenko1976,Ballou1986,Katayama1976}/YCo$_5$~\cite{Alameda1981,Yermolenko1980}.

Up to 200~K, there is close agreement between the experiments of
Ref.~\cite{Ermolenko1976}, our own experiments, and the 
\FPMvB~calculations.
Above this temperature our new experiments show the expected
drop in $K_\mathrm{eff}$, while the previously reported
data show a continued rise~\cite{Ermolenko1976}.
We repeated our measurements using different protocols and
found a reasonably large variation in the extracted $K_\mathrm{eff}$~\cite{SM}.
Even taking this variation into account as the shaded area in Fig.~\ref{fig.KvT},
the drop is still observed.

We therefore do not believe the high temperature behavior reported
in Ref.~\cite{Ermolenko1976} has an intrinsic origin.
Possible extrinsic factors include the method of sample preparation,
degradation of the RCo$_5$ phase 
at elevated temperatures~\cite{Denbroeder1972}, and potential systematic
error when extracting $K_\mathrm{eff}$.
We note that even the idealized theoretical 
curves in Fig.~\ref{fig.MvB} show curvature at higher
temperature, making it more difficult to find the intercept.

In conclusion, we have introduced the \FPMvB \ approach
to interpret experiments measuring anisotropy of ferrimagnets,
particularly RE-TM permanent magnets.
We presented the method in the context of our DLM formalism,
but any electronic structure theory capable of calculating
magnetic couplings relativistically~\cite{Udvardi2003,Ebert2009,Hu2013,Khan20162,Hoffmann2017} 
should be able to produce \FPMvB \ 
curves, at least at zero temperature.
However standard calculations which neglect the external field should be used
with care when comparing to experiments on ferrimagnets.
Similarly, the prototype GdCo$_5$ 
serves as a reminder that a simple view of the anisotropy 
energy does not fully describe the magnetization 
processes in ferrimagnets, which might have implications
in understanding e.g.\  magnetization
reversal in nano-magnetic assemblies~\cite{Guo2002}.
Overall our work demonstrates the benefit of 
interconnected computational
and experimental research in this key area.

The present work forms part of the PRETAMAG project, 
funded by the UK Engineering and  Physical  Sciences 
Research  Council (EPSRC),  Grant  no.\   EP/M028941/1.
Crystal growth work at Warwick is also supported by 
EPSRC Grant no.\ EP/M028771/1.
Work  at  Daresbury Laboratory was supported by an EPSRC service 
level agreement with the Scientific Computing Department of STFC. 
We thank E.\ Mendive-Tapia  for useful discussions and 
A.\ Vasylenko for continued assistance in translating
references.


%
\end{document}